\documentclass[journal, 12pt, onecolumn, draftclsnofoot]{IEEEtran}
\usepackage{authblk}
\usepackage[utf8]{inputenc}
\usepackage{mathtools}
\usepackage{graphicx}
\usepackage[table]{xcolor}
\usepackage{xcolor}
\usepackage{amsthm}
\usepackage{algorithm}
\usepackage{algpseudocode}
\usepackage{amssymb}
\usepackage{cite}
\usepackage{multirow}
\usepackage{bm}
\usepackage{tablefootnote}
\usepackage{threeparttable}
\usepackage{braket}
\usepackage{booktabs}
\usepackage{hyperref}
\usepackage{caption}
\usepackage{hhline}

\usepackage{tikz}
\usepackage{tikzpeople}
\usetikzlibrary{tikzmark, shapes, fit,backgrounds}
\usetikzlibrary{matrix, decorations.pathreplacing, angles, quotes}
\usetikzlibrary{patterns, calc, positioning}
\usetikzlibrary{arrows, arrows.meta}
\usetikzlibrary{shapes.multipart}
\usetikzlibrary{fit}
\usetikzlibrary{shapes.geometric, positioning}
\usetikzlibrary{decorations.pathmorphing}

\tikzset{meter/.append style={draw, inner sep=10, rectangle, font=\vphantom{A}, minimum width=30, scale=.7, path picture={\draw[black] ([shift={(.1,.3)}]path picture bounding box.south west) to[bend left=50] ([shift={(-.1,.3)}]path picture bounding box.south east);\draw[black,-{Latex[scale=.5]}] ([shift={(0,.1)}]path picture bounding box.south) -- ([shift={(.3,-.1)}]path picture bounding box.north);}}}
\tikzset{snake it/.style={decorate, decoration=snake}}

\newtheorem{theorem}{Theorem}

\newtheorem{remark}{Remark}

\theoremstyle{definition}

\newcommand{\init}{{\scalebox{0.7}{\mbox{init}}}}

\allowdisplaybreaks
\begin{document}
\title{\Large On the Capacity of Secure $K$-user Product Computation over a Quantum MAC}
\author{\normalsize Yuxiang Lu, Yuhang Yao and Syed A. Jafar}
\affil{\small Center for Pervasive Communications and Computing (CPCC), UC Irvine\\
Email: \{yuxiang.lu, yuhangy5, syed\}@uci.edu}
\date{}
\maketitle
\begin{abstract}
 Inspired by recent work by Christensen and Popovski on secure $2$-user product computation for finite-fields of prime-order over a quantum multiple access channel, the generalization to $K$ users and arbitrary finite fields is explored. Asymptotically optimal (capacity-achieving for large alphabet) schemes are proposed. Additionally, the capacity of modulo-$d$ ($d\geq 2$) secure $K$-sum computation  is shown to be $2/K$ computations/qudit, generalizing a result of Nishimura  and Kawachi beyond binary, and improving upon it for odd $K$.
\end{abstract}

\section{Introduction}
Secure multiparty quantum protocols for fundamental  primitives  such as summation and multiplication  have been explored under a variety of idealized models \cite{QSumProd, PSQM, christensen2023private, Yao_Jafar_SumMAC_Quantum, song2019capacitycollusion, song2020capacity,PSQMsurvey}. We focus in particular, on  the \emph{private simultaneous quantum messages} (PSQM) setting introduced in 2021 by Nishimura and Kawachi \cite{PSQM}. Related multiparty computation models are surveyed in \cite{PSQMsurvey}. The PSQM setting involves $K$ users with  private data $W_1,\cdots, W_K$, and a server who computes a function $F(W_1,\cdots,W_K)$, without learning anything else about the users' inputs. Common randomness $(Z)$ and quantum entanglement $(\mathcal{Q})$ are distributed to  the $K$ users.  Each user manipulates its quantum-subsystem locally and sends it to the server, who then recovers $F$ with zero error. Reference \cite{PSQM} explores the communication \emph{complexity}, i.e., the number of qubits of communication needed to compute one instance of $F$. In this work, however, we take a perspective common in information theory and focus instead on the  computation \emph{rate}, i.e., the number of instances of $F$ that can be computed per qubit of communication cost. Note that this allows for batch-processing, i.e.,  amortization of cost by joint computation of multiple instances of $F$. The fundamental limit of the computation rate is  the information-theoretic \emph{capacity}.

{\it $\hspace{0.1cm}$ Motivation:} Some motivating questions for this work are listed next, labeled as Q1-Q5 for reference.  Reference \cite{PSQM} explores various Boolean functions  and presents in \cite[Lemma 10]{PSQM} a PSQM scheme that computes a $K$-user modulo-$2$ sum, achieving a rate $2/K$ computations/qubit if $K$ is even and a rate $2/(K+1)$ if $K$ is odd. In \cite[Lemma 11]{PSQM} another scheme is presented that computes a $K$-user generalized equality  function {\it GEQ$_{n}$} (each user has an $n$ bit input vector, $n$ is even, and the goal is to determine if the bit-wise modulo-2 sum of the input vectors is the all-zero vector), that achieves the rate $2/(Kn)$ if $K$ is even and $2/((K+1)n)$ if $K$ is odd. For both schemes, the possibility of rate improvements for odd $K,n$ is open (Q1). Based on these schemes, it follows from \cite[Theorem 2]{PSQM}  that there exists a total function $F: (\{0,1\}^{n})^K\rightarrow\{0,1\}$ (each user has $n$ bit input, output is $1$ bit) for which quantum entanglement improves rate by a factor of $2$ \emph{if $K,n$ are both even}. It is not known (Q2) if the claim extends to odd $K,n$.  Another important development is the recent work by Christensen and Popovski, who propose in \cite{christensen2023private} PSQM  secure product computation  schemes for $2$ users over any prime field $\mathbb{F}_d$. The possibility of rate improvements is  open (Q3). Other open problems  in \cite{christensen2023private} include  generalizations to  finite fields $\mathbb{F}_d$ for non-prime $d$ (Q4), and from $2$-users  to $K$-users (Q5).

\begin{figure}[h]
    \centering
    \includegraphics{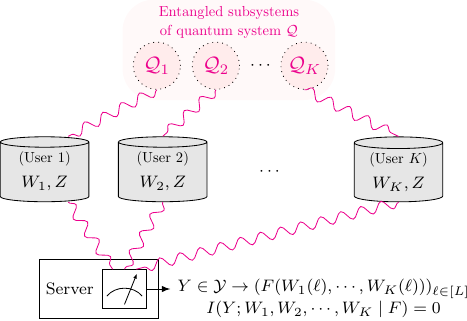}
    \caption{Secure $K$-user Computation over a Quantum MAC}\label{fig:KPC}
\end{figure}

{\it $\hspace{0.1cm}$ Contribution:} We prove that the capacity of secure $K$-user  modulo-$d$ sum computation is  $2/K$ computations/qudit. Besides the generalization from binary to modulo-$d$ addition, the new scheme strictly improves the rate  from $2/(K+1)$  in \cite[Lemma 10]{PSQM} to $2/K$ computations/qubit when $K$ is odd. Using the capacity achieving scheme and batch processing in  \cite[Lemma 11]{PSQM} achieves the rate of $2/(Kn)$ for {\it GEQ$_{n}$} for \emph{all} $K,n$, thereby answering Q1. Furthermore, using the improved rate {\it GEQ$_{n}$} scheme in \cite[Theorem 2]{PSQM} shows that there exists a total function $F: (\{0,1\}^{n})^K\rightarrow\{0,1\}$ for which quantum entanglement improves computation rate by a factor of $2$, for \emph{all}  $K,n$, thus answering Q2.

For the secure product computation problem of \cite{christensen2023private}, as answers to Q4 and Q5, we present a  $K$-user product  computation scheme over any finite field $\mathbb{F}_d$ ($d$ need not be prime), achieving the  rate of at least $(2/K)/[\log_d(2K-1)+\log_d(d-1)]$, which is asymptotically optimal (asymptotic capacity is $2/K$ computations/qudit) for large alphabet $d$. In addition to batch-processing, our scheme combines ideas from the $2$-sum protocol \cite{song2019capacitycollusion, song2020capacity}, additive secret sharing, the FKN scheme \cite{FKN}, and a field-group isomorphism \cite{jia2021SDBMM}. 

In terms of Q3, for the $2$-user secure product over $\mathbb{F}_2$ (equivalently, the secure AND computation), the rate is improved from $1/2$ in \cite{christensen2023private} to $1/\log_2(3)$ computations/qubit. For large alphabet (i.e., $2$-user secure product over $\mathbb{F}_d$ for large prime-power $d$) the improvement approaches a factor of $2$.

\begin{figure}[htbp]
    \centering
    \includegraphics[width=0.48\textwidth]{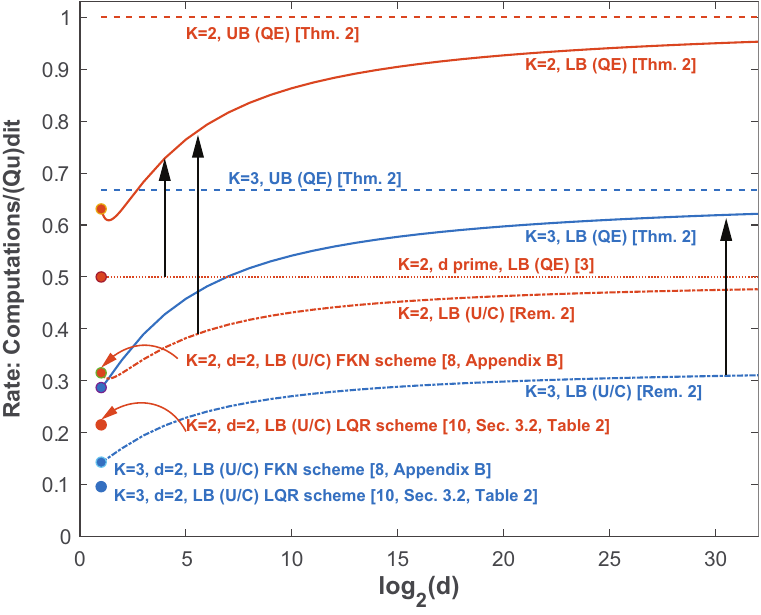}
    \caption{\small Rates for QSK-Prod (UB: Upper bound, LB: Lower bound, i.e., achievable rate, QE: with Quantum Entanglement, U/C: Unentangled/Classical Setting, Rem2 and Thm2: Remark 2 and Theorem 2 in this letter). Thick circles mark achievable rates for $d=2$ (AND). }\label{fig:numerical}
\end{figure}

For further insights consider classical schemes for secure product computation, namely the FKN scheme \cite[Appendix B]{FKN} and the  Linear Quadratic Residue scheme (LQR) of \cite[Table 2]{QRPSM}, which is limited to $d=2$, i.e., the AND function. A classical scheme for arbitrary $\mathbb{F}_d$ is devised in Remark 2 in this work and shown in Fig. \ref{fig:numerical} as a baseline for comparison. Evidently, our secure-product schemes   improve significantly on the quantum baseline from \cite{christensen2023private} as well as the classical baselines, as shown by  vertical arrows in Fig. \ref{fig:numerical}.

Table \ref{tab:QSKAND} shows quantum-advantage vs security-penalty,  listing the best-known achievable costs (qubits/computation, reciprocal of rate) for $K$-user AND. $I_K^o$ is   $1$ if $K$ is odd and $0$ otherwise. The classical cost without security is $K$, as all $K$ inputs are necessary to compute the AND. For quantum cost without security, the best scheme we are able to devise  groups the users into $\lfloor K/2\rfloor$ pairs and computes the AND  for each pair with the scheme from Section \ref{sec:qs2and} at the cost of $\log_2(3)$ qubits/computation. When $K$ is odd, the remaining user  sends its input via a qubit. The product of pairwise ANDs and the remaining user's input yields the overall AND.
 
\begin{table}[h]
\caption{Quantum/Classical and Secure/Insecure $K$-user AND}
\label{tab:QSKAND}
\centering
\begin{tabular}{|l|c|l|}
\hline
\cellcolor{black!10} & \cellcolor{black!10}  & \cellcolor{black!10} \\[-0.2cm]
\cellcolor{black!10}Cost & \cellcolor{black!10}  With security & \cellcolor{black!10} Without security\\[0.1cm]\hline
&&\\[-0.2cm]
Quantum & $(K\log_2(2K-1))/2$ [Thm2] & $\lfloor K/2\rfloor\log_2(3)+I_K^o$ \\[0.1cm]\hline
&&\\[-0.2cm]
Classical & $K\log_2(2K-1)$  [Rem2]& $K$ (trivial)\\[0.1cm]\hline
\end{tabular}
\end{table}

{\it Notation:} For integers $a \leq b$,  $[a:b]\triangleq\{a,a+1,\cdots,b\}$,  $Y_{[a:b]}\triangleq\{Y_a, Y_{a+1}, \cdots, Y_{b}\}$ and $[b]\triangleq[1:b]$. The ring of integers modulo $d$ is $\mathbb{Z}_d = \mathbb{Z} / d\mathbb{Z}$.  $A \land B$ is the logical AND of binary $A,B$. Define $1(x)=1$ if $x\neq 0$ and $0$ otherwise.

\section{Problem Statement}\label{sec:form}
As shown in Fig. \ref{fig:KPC}, there is one server, $K$ users, and $K$ private data streams, such that the $k^{th}$ private data stream $W_k$ is available only to User $k$, $k\in[K]$. $W_k(\ell) \in \mathcal{W}$ denotes the $\ell^{th}$ instance of $W_k$. The function to be securely computed  is  $F \colon \mathcal{W}^K \rightarrow \mathcal{F}$ for some output alphabet $\mathcal{F}$, and $F(W_1(\ell), \cdots, W_K(\ell))\triangleq F(\ell)$  is the $\ell^{th}$ instance of the desired secure computation.  
Three types of  Quantum Secure $K$-user Computation (QSK-Comp) settings are considered, as defined in Table \ref{tab:4QSKComp}.  QSK-AND is same as QSK-Prod for $d=2$. 
\begin{table}[h]
\caption{Three types of QSK-Comp Settings}
\label{tab:4QSKComp}
\centering
\begin{tabular}{|l|c|l|}
\hline
\cellcolor{black!10} & \cellcolor{black!10}  & \cellcolor{black!10} \\[-0.2cm]
\cellcolor{black!10}QSK-Comp & \cellcolor{black!10}  Alphabet $(\mathcal{W},\mathcal{F})$ $(d)$ & \cellcolor{black!10} Function $F$\\[0.1cm]\hline
&&\\[-0.2cm]
QSK-AND  & $(\mathbb{F}_2,\mathbb{F}_2)$ $(d=2)$ & $F = W_1 \land W_2 \land \cdots \land W_K$\\[0.1cm]\hline
&&\\[-0.2cm]
QSK-Sum  & $(\mathbb{Z}_d,\mathbb{Z}_d)$ ($d \geq 2$)  & $F = W_1 + W_2 + \cdots + W_K$\\[0.1cm]\hline
&&\\[-0.2cm]
QSK-Prod  & $(\mathbb{F}_d,\mathbb{F}_d)$ ($d = p^r$) & $F = W_1 W_2 \cdots W_K$ \\[0.1cm]\hline
\end{tabular}
\end{table}

A QSK-Comp coding scheme is specified by a $7$-tuple $(L,Z, \delta_{[K]}, \rho_{\init},\Phi_{[K]},\{M_y\}_{y\in\mathcal{Y}}, \Psi)$. 
The batch size $L\in \mathbb{N}$ is the number of data instances to be encoded together. 
$Z$ is the common randomness, independent of the data streams. 
For $k\in [K]$, denote $W_k^{(L)} = [W_k(1),\cdots,W_k(L)]^T$, and $F^{(L)} = [F(1),\cdots,F(L)]^T$. 
 The composite quantum system $\mathcal{Q}$  is initially described by its density matrix $\rho_{\init} \in \mathbb{C}^{\delta\times \delta}, \delta\triangleq \delta_1\delta_2\cdots \delta_K$, independent of $(W_1,\cdots,W_K,Z)$. User $k$ is allocated the $\delta_k$-dimensional quantum subsystem $\mathcal{Q}_k$. 
For $k\in [K]$, $\Phi_k(W_k^{(L)},Z) = U_k$ is a unitary operator which is applied to $\mathcal{Q}_k$. The resulting state of the overall quantum system is $\rho = U \rho_{\init}U^\dagger$, where $U=U_1\otimes \cdots \otimes U_K$. 
The server applies POVM $\{M_y\}_{y\in \mathcal{Y}}$ and gets $Y\in \mathcal{Y}$ as the output. Finally, $\Phi(Y) = F^{(L)}$ recovers the desired computation. The scheme must correctly recover $F^{(L)}$ for every realization of $(W_1^{(L)},\cdots,W_K^{(L)})$.
Also, the scheme must be  secure, i.e., for any data realizations $(W_1^{(L)},\cdots,W_K^{(L)}) = (w_1,\cdots, w_K)$ and $(w_1',\cdots, w_K')$ that yield the same $F^{(L)}$, we require that $\rho$ and $\rho'$ have the same distribution, where $\rho$ and $\rho'$ are the corresponding received states, respectively.

A rate $R$ is feasible if there exists a secure quantum coding scheme $$(L,Z, \delta_{[K]}, \rho_{\init},\Phi_{[K]},\{M_y\}_{y\in\mathcal{Y}}, \Psi)$$ so that $R\leq L/\log_d(\delta)$. The unit of  rate is computations/qudit (i.e., the number of instances of $d$-ary $F$ computed per $d$-dimensional quantum system download), where $d$ is as specified in Table \ref{tab:4QSKComp}. The capacity $C$ is  the supremum of all feasible rates.

\section{Results}

\subsection{Pre-requisite: Modulo-$d$ $2$-sum protocol of \cite{song2020capacity,song2019capacitycollusion}}
Consider two transmitters, Alice and Bob, with inputs $A_1, A_2\in\mathbb{Z}_d$ available to Alice and $B_1, B_2\in\mathbb{Z}_d$ available to Bob. Alice and Bob possess one each of a pair of  qudits in an entangled state, namely the  Bell state $\ket{\phi^{0,0}}_{AB} \triangleq \frac{1}{\sqrt{d}}\sum_{i=0}^{d-1}\ket{i}_A\ket{i}_B$. Define  $\ket{\phi^{x,z}}_{AB} \triangleq ({\sf X}_A^x{\sf Z}_A^z\otimes \mathbf{I}_B) \ket{\phi^{0,0}}_{AB}$. According to \cite[Proposition III.1]{song2020capacity},  $\{\ket{\phi^{x,z}}_{AB} \mid x, z \in \mathbb{Z}_d\}$ forms an orthonormal basis, so measurement in this basis determines $x,z$. Alice and Bob apply ${\sf X}$ and ${\sf Z}$ gates according to their data, resulting in the state $({\sf X}_A^{A_1}{\sf Z}_A^{A_2}\otimes {\sf X}_B^{-B_1}{\sf Z}_B^{B_2}) \ket{\phi^{0,0}}_{AB}=\ket{\phi^{x,z}}_{AB}$ where $x = A_1 + B_1, z = A_2 + B_2$, and the addition is modulo $d$.  Note that the state of the quantum system is fully determined by $(x,z)$, i.e., it can reveal nothing besides $A_1+B_1, A_2+B_2$.  

\begin{table}[htbp]
\caption{The Modulo-$d$ $2$-sum Protocol \cite{song2020capacity, song2019capacitycollusion}.}\label{tab:two_sum}
\center
\begin{tabular}{c|c|c|c}
\toprule
	Input (Alice) & Input (Bob) & Output (Charlie) & Cost \\\hline
    $(A_1,A_2) \in \mathbb{Z}_d^2$  &  $(B_1,B_2) \in \mathbb{Z}_d^2$ & $(A_1+B_1,A_2+B_2)$ & $2$ qudits \\
    \bottomrule
\end{tabular}
\end{table}

\noindent As shown in Table \ref{tab:two_sum}, since $2$ instances of the sums are computed with the total communication cost of $2$ qudits, one from each transmitter, the normalized cost  is $1$ qudit/sum. Therefore, in our batched-setting the two-sum protocol may be equivalently viewed as a classical modulo-$d$ additive channel, with the cost of one qudit  per channel use, as in Table \ref{tab:addchannel}. Note that without quantum entanglement  the minimum communication cost even without security constraints is at least \emph{two} qudits/sum, thus demonstrating the advantage made possible by quantum entanglement.
\begin{table}[htbp]
\caption{Equivalent channel representation of the $2$-sum Protocol.}
\label{tab:addchannel}
\center
\begin{tabular}{c|c|c|c}
\toprule
	Input (Alice) & Input (Bob) & Output (Charlie) & Cost \\\hline
    $A \in \mathbb{Z}_d$  &  $B \in \mathbb{Z}_d$ & $(A+B)\in\mathbb{Z}_d$ & $1$ qudit \\
    \bottomrule
\end{tabular}
\end{table}

\subsection{QS2-AND}\label{sec:qs2and}
Let us  show how the QS2-AND protocol of \cite{christensen2023private} is also achieved via the  $2$-sum protocol. Let $A,B\in \mathbb{Z}_2$ be the inputs available to Alice and Bob, respectively. Let $Z$ be a random variable uniformly drawn from $\{1,2,3\}$, which is shared only between Alice and Bob. According to the realizations of $Z$, Alice and Bob use the two-sum protocol as shown in  Table \ref{tab:protocol_CP}.
\begin{table}[htbp]
\caption{QS2-AND Protocol in \cite{christensen2023private}.}
\label{tab:protocol_CP}
\center
\begin{tabular}{c|c|c|c}
\toprule
	 & $Z=1$ & $Z=2$ & $Z=3$ \\\hline
    Input (Alice)  &  $(A,0)$ & $(0,A)$  & $(A+1,A)$  \\\hline
    Input (Bob)  &  $(0,B)$ & $(B,B+1)$  & $(B,0)$  \\\hline
    Output (Charlie)  & $(A,B)$  & $(B,A+B+1)$  & $(A+B+1,A)$ \\ \bottomrule
\end{tabular}
\end{table}
The scheme is correct because under all realizations of $Z$, the output at Charlie is equal to $(1,1)$ if $A=B=1$, and uniform over $\{(0,0), (0,1),(1,0)\}$ otherwise, which also guarantees security. Since this scheme securely computes $1$ instance of the AND function, with the total communication cost of $2$ qubits, it achieves rate $0.5$ (computations/qubit), and requires $\log_2(3)$ bits of  common randomness (i.e., a uniform $3$-ary $Z$) per computation.

As our first result let us present a scheme for QS2-AND,  that achieves a higher rate: $1/\log_2 (3) \approx 0.63$ (computations/qubit) instead of $0.5$, while also requiring less classical common randomness: $1$ bit instead of $\log_2(3)$ bits per computation. For the new scheme we combine the additive channel of Table \ref{tab:addchannel} with the FKN scheme \cite[Appendix B]{FKN}. The FKN scheme is a classical scheme which enables the server to securely compute the AND function once, with each user sending $\log_2 (3)$ bits to the server. The rate achieved by the FKN scheme is $1/\big(2\log_2 (3)\big)$ (computations/bit), which is optimal for the classical setting as shown by \cite{Data_Prabhakaran_Prabhakaran}. To construct our QS2-AND protocol based on the FKN scheme, specifically, we let $d=3$ and use common randomness $Z$ that is uniform in $\{1,2\}$. The new protocol is shown in Table \ref{tab:QS2Prod_proposed}.
\begin{table}[htbp]
\caption{New QS2-AND Protocol}
\label{tab:QS2Prod_proposed}
\center
\begin{tabular}{c|c|c}
\toprule
	 & $Z=1$ & $Z=2$ \\\hline
    Input (Alice)  &  $(1-A)$ & $2(1-A)$  \\\hline
    Input (Bob)  &  $(1-B)$ & $2(1-B)$ \\\hline
    Output (Charlie)  & $(1-A)+(1-B)$  & $2((1-A)+(1-B))$ \\ \bottomrule
\end{tabular}
\end{table}
For the correctness of this scheme note that the output at Charlie is equal to $0$ if and only if  $A\land B = 1$.  Security  is also guaranteed as given any $(A,B) \neq (1,1)$, the distribution of the output at Charlie is the same (uniform over $\{1, 2\}$).
The rate achieved is $1/\log_2 (3)$ (computations/qubit). Note that in the classical setting the  FKN scheme in \cite{FKN} also  requires an additional $\log_2 (3)$ bits of common randomness that serves as additive noise which is not needed in the QS2-Prod scheme, because of the inherent additive property of the two-sum protocol over the quantum MAC. As a result, the FKN scheme needs $1+\log_2 (3) = \log_2 (6)$ bits of common randomness per computation, while our QS2-AND only needs $1$ bit of common randomness per computation.  Relative to the unentangled/classical setting represented by the FKN scheme, the quantum entanglement advantage is $2$-fold,  reflected  in both higher rate and lower requirement of common randomness.

The generalization to QSK-AND appears in Section \ref{sec:QSKprod} as a special case of QSK-Prod. 

\subsection{QSK-Sum}\label{sec:QSKsum}
The capacity for QSK-Sum (modulo-$d$) is stated below. 

\begin{theorem} \label{thm:QSKSum}
	The capacity of QSK-Sum is $C_{s} = 2/K$.
\end{theorem}
{\it Proof: } Let us start with the proof of achievability.
\subsubsection{Achievability}
The case of even $K$ is simple. For $d=2$ (binary sums) the achievability of rate $2/K$ already follows from \cite[Lemma 10]{PSQM}. The generalization to arbitrary $d$ uses additive secret sharing and the $2$-sum protocol. Specifically,  $\forall i\in [K/2]$, User $2i-1$ and User $2i$ use the additive channel of Table \ref{tab:addchannel} to transmit the modulo-$d$ sum  $W_{2i-1}+W_{2i}+Z_i$, where $Z_2, Z_3,\cdots, Z_{K/2}$ are i.i.d. uniform in $\mathbb{Z}_d$  and  $Z_1+Z_2+\cdots+Z_{K/2} = 0$. The server computes the modulo-$d$ sum $\sum_{k=1}^K W_{k} + \sum_{i=1}^{K/2}Z_i = \sum_{k=1}^K W_{k}$ by adding (modulo-$d$) the  sums received from the $K/2$ pairs of  users.

Now say $K$ is odd. Since $K=1$ is trivial, assume $K\geq 3$. Consider $L=2$  data instances, i.e., $W_k(1)$ and $W_k(2)$ for each data stream. Let Users $1$, $2$, Users $1$, $3$ and Users $2$, $3$ use the additive channel once each (thus three times among the first three users). For the remaining users, Users $2i$, $2i+1$ use the additive channel twice for each $i\in \{2,3,\cdots, (K-1)/2\}$. The inputs and outputs of the additive channels are specified in Table \ref{tab:oddKscheme}, where $Z_0, Z_i(\ell),\forall i\in[2:(K-1)/2], \ell \in [2]$ are i.i.d. uniform in $\mathbb{Z}_d$ and
\begin{align}
Z_1(\ell)+Z_2(\ell)+\cdots+Z_{(K-1)/2}(\ell) = 0, &&\forall \ell \in\{1,2\}.\label{eq:sumzero}
\end{align} 

\begin{table}[htbp]
\caption{QSK-Sum (modulo-$d$) scheme for odd $K$.}
\label{tab:oddKscheme}
\center
\scalebox{0.84}{
\begin{tabular}{@{\hskip -0.05cm}r@{\hskip -0.02cm}|l}
\toprule
\begin{tabular}{lr}User index:& ~~~~~~~~~Input\end{tabular}~	& ~~Output $=$ Sum of the two inputs  \\[0.1cm]\hline
&\\[-0.2cm]
\begin{tabular}{lr}
    1:& $W_1(1)-W_1(2)$\\[0.03cm]
    2:&  $W_2(1)+Z_0+Z_1(1)$
\end{tabular} & 
$\begin{array}{l}
Y_1= W_1(1)-W_1(2)+W_2(1)+Z_0+Z_1(1)
\end{array}$
\\[0.3cm]\hline
&\\[-0.2cm]
\begin{tabular}{lr}
    2:& $W_2(2)+Z_0+Z_1(2)$\\[0.03cm]
    3:&  $W_3(2)-W_3(1)$
\end{tabular} & 
$\begin{array}{l}
Y_2 = W_2(2)-W_3(1)+W_3(2)+Z_0+Z_1(2)
\end{array}$
\\[0.3cm]
\hline
&\\[-0.2cm]
\begin{tabular}{lr}
    1:& \hspace{1cm}$W_1(2)$\\[0.03cm]
    3:& \hspace{1cm} $W_3(1)-Z_0$
\end{tabular} & 
$\begin{array}{l}
Y_3 = W_1(2)+W_3(1)-Z_0
\end{array}$
\\[0.3cm]
\hline
&\\[-0.2cm]
\begin{tabular}{lr}
    $2i$:& $W_{2i}(1)+Z_i(1)$\\[0.03cm]
    $2i+1$:&  $W_{2i+1}(1)$
\end{tabular} & 
$\begin{array}{r}
Y_{2i}=W_{2i}(1)+W_{2i+1}(1)+Z_i(1)\\[0.1cm]
\scalebox{0.9}{$2\leq i\leq (K-1)/2$}
\end{array}$
\\[0.3cm]
\hline
&\\[-0.2cm]
\begin{tabular}{lr}
    $2i$:& $W_{2i}(2)+Z_i(2)$\\[0.03cm]
    $2i+1$:&  $W_{2i+1}(2)$
\end{tabular} 
& 
$\begin{array}{r}
Y_{2i+1} = W_{2i}(2)+W_{2i+1}(2)+Z_i(2)\\[0.1cm]
\scalebox{0.9}{$2\leq i\leq (K-1)/2$}
\end{array}$
\\[0.3cm] 
    \bottomrule
\end{tabular}
}
\end{table}
In Table \ref{tab:oddKscheme}, each $Y_i$ represents the output for one use of the additive channel of Table \ref{tab:addchannel}, e.g., the first row of Table \ref{tab:oddKscheme} means that User $1$ and User $2$ use the additive channel once, with User $1$'s input specified as $W_1(1)-W_1(2)$ and User $2$'s input specified as $W_2(1)+Z_0+Z_1(1)$. The output is $Y_1 = W_1(1)-W_1(2)+W_2(1)+Z_0+Z_1(1)$.

\noindent {\bf Correctness:} The server is able to recover the two computation instances $F(1)$ and $F(2)$ as,
\begin{align}
&\underbrace{Y_1+Y_3}_{W_1(1)+W_2(1)+W_3(1)+Z_1(1)}+\underbrace{Y_4+Y_6+\cdots+Y_{K-1}}_{\sum_{i=2}^{(K-1)/2}W_{2i}(1)+W_{2i+1}(1)+Z_i(1)}\notag\\
&=W_1(1)+W_2(1)+\cdots+W_K(1) = F(1),\label{eq:recf1}\\
&\mbox{and similarly, } Y_2+Y_3+Y_{5}+Y_{7}+\cdots+Y_{K}=F(2).\label{eq:recf2}
\end{align}%
\noindent {\bf Security:} The  scheme is based on the two-sum protocol. Recall that the received quantum states are deterministic (pure states) conditioned on the measurement outputs. It therefore suffices to show that all the outputs of the two-sum protocols are collectively  independent of the data, conditioned on the desired computation $F$. Let us prove this for the case where $K$ is odd. The proof for even $K$ follows similarly.
\begin{align}
&I(W_{[K]}^{(2)}; Y_{[K]}\mid F^{(2)})\notag\\
&=H(Y_{[K]}\mid F^{(2)})-H(Y_{[K]}\mid  F^{(2)}, W_{[K]}^{(2)})\label{eq:step1}\\
&=H(Y_{[K]}\mid F^{(2)})-H(Z_0,Z_{[(K-1)/2]}^{(2)}\mid  F^{(2)}, W_{[K]}^{(2)})\label{eq:step3}\\
&=H(Y_{[K]}\mid F^{(2)})-H(Z_0,Z_{[(K-1)/2]}^{(2)})\label{eq:step4}\\
&=H(Y_{[K]}\mid F^{(2)})-H(Z_0,Z_{[2:(K-1)/2]}^{(2)})\label{eq:step5}\\
&=H(Y_{[K]}\mid F^{(2)})-\left(K-2\right)\log(d)\label{eq:step7}\\
&=H(Y_{[3:K]}\mid F^{(2)})-\left(K-2\right)\log(d)\label{eq:step8}\\
&\leq H(Y_{[3:K]})-\left(K-2\right)\log(d)\label{eq:step9}\\
&\leq 0\label{eq:step10}
\end{align}
Step \eqref{eq:step3} holds because given all data streams $(W_{[K]}^{(2)})$, the tuple containing all common randomness terms $(Z_0, Z_{[(K-1)/2]}^{(2)})$ is an invertible function of  $Y_{[K]}$ (see Table \ref{tab:oddKscheme}). Step \eqref{eq:step4} similarly holds because the common randomness is independent of the data and the function to be computed. Step \ref{eq:step5} holds because $Z_1(\ell)$ is determined by $Z_2(\ell),Z_3(\ell),\cdots,Z_{(K-1)/2}(\ell)$ for $\ell=1,2$ according to \eqref{eq:sumzero}. Step \ref{eq:step7} holds because the remaining common randomness terms are i.i.d. uniform in $\mathbb{Z}_d$ and there are  $K-2$ of them. 
Step \ref{eq:step8} used the fact that $Y_{[K]}$ is an invertible function of $(F(1), F(2), Y_3,Y_4,\cdots,Y_K)$ according to \eqref{eq:recf1}, \eqref{eq:recf2}. Step \eqref{eq:step9} follows because conditioning reduces entropy, and Step \eqref{eq:step10} uses the fact that uniform distribution maximizes entropy. Thus, the derivation shows that $I(W_{[K]}^{(2)}; Y_{[K]}\mid F^{(2)})\leq 0$. Since mutual information cannot be negative we must have $I(W_{[K]}^{(2)}; Y_{[K]}\mid F^{(2)})= 0$. Thus the protocol is secure.

\noindent {\bf Rate:} For even $K$, the additive channel is used $K/2$ times in order to compute the $K$-sum \emph{once}. For odd $K$, the additive channel is used a total of $K$ times in order to compute the $K$-sum \emph{twice}. Thus, the rate achieved in both cases is $2/K$.

\subsubsection{Converse} \label{proof:converse_QSKSum}
Let us show that even without the security constraint the rate achieved  cannot be more than $2/K$. 
Suppose there exists a feasible coding scheme $\mathcal{C}$ for the tuple\\ $(L,Z, \delta_{[K]}, \rho_{\init},\Phi_{[K]},\{M_y\}_{y\in\mathcal{Y}}, \Psi)$. Let us show that  $2\log_d (\delta_k) \geq L$  for \emph{every} $k\in[K]$. Start with $k=1$. Note that a feasible coding scheme must allow correct decoding for \emph{all} data realizations. So let User $1$'s data $W_1^{(L)}$ be uniformly distributed in $\mathbb{F}_d^{L}$,  while all other users' data is constant, say all zeros. Let Users $2,3,\cdots, K$ and the server combine all their resources, forming, say a super-server. Because the coding scheme $\mathcal{C}$ is correct by assumption, applying the scheme must allow the super-server  to recover $F^{(L)}$, and therefore recover $W_1^{(L)}$ because $W_1^{(L)}=F^{(L)}$ when the other users' data is all zeros. The super-server and User $1$ now share quantum entanglement, but since the only communication between them is the $\delta_1$ dimensional quantum system $\mathcal{Q}_1$, it follows from the \emph{information causality bound} \cite[Prop. 6]{massar2015hyperdense}, \cite[IV.A]{Yao_Jafar_SumMAC_Quantum}  that  $2\log_d (\delta_1) \geq I(W_1^{(L)};Y) =H(W_1^{(L)})= L$, where the mutual information is measured in dits ($\log_d(\cdot)$). Repeating the same argument for $k=2,3,\cdots, K$ we have $2\log_d (\delta_k)\geq L$ for all $k\in[K]$. Thus, $2\log_d(\delta_1\delta_2\cdots \delta_K) \geq KL \implies L/\log_d(\delta_1\delta_2\cdots\delta_K) \leq 2/K$ and thus $C\leq 2/K$. $\hfill\blacksquare$

\begin{remark} Note that the achievable scheme requires only additive inverses (which exist over both rings and finite fields), and the converse applies over finite fields as well. It follows that the capacity of QSK-Sum over finite fields is also $2/K$, extending the corresponding result of \cite{Yao_Jafar_SumMAC_Quantum} to the secure setting.
\end{remark}

\subsection{QSK-Prod}\label{sec:QSKprod}

\begin{theorem} \label{thm:qskprod}
	The capacity of QSK-Prod is bounded as follows.
	\begin{align}
		 \frac{2/K}{\log_d(2K-1)+\log_d(d-1)}\leq C_p\leq 2/K.
	\end{align}
	In particular, as $d\to\infty$ the asymptotic capacity is $2/K$. 
\end{theorem}
{\it Proof:} We begin with the proof of achievability.
\subsubsection{Achievability}
The achievable scheme consists of two phases. The first phase allows the server to securely compute $1(W_1)1(W_2)\cdots 1(W_K)$, which is a QSK-AND problem. If $d=2$ then the computation is complete, otherwise the second phase allows the server to securely compute the product $W_1W_2\cdots W_K$ if the AND computation is non-zero.\footnote{$W_k$ here denotes one instance of the $k^{th}$ data stream.} Both phases make use of the QSK-Sum scheme, and for $d>2$ both phases are performed in all cases to preserve security.

\noindent {\bf Phase I (QSK-AND):} In the first phase, the server securely computes $1(W_1)1(W_2)\cdots 1(W_K)$, combining the $2$-sum protocol and the FKN scheme \cite[Appendix B]{FKN}. To do this, let $p$ be a prime such that $K<p<2K$. The existence of such a $p$ is guaranteed by the  Bertrand–Chebyshev Theorem \cite{chebyshev1852memoire}. Then let $R$ be uniformly distributed in $\mathbb{F}_p \backslash \{0\}$. We apply the QSK-Sum scheme over $\mathbb{F}_p$ and let the input from User $k$ be $R(1-1(W_k))$, for all $k\in [K]$. The QSK-Sum scheme allows the server to securely compute $Y = R\sum_{i=1}^K (1-1(W_k))$, with a cost $(K/2) \log_d(p)$ qudits/computation. Note that $Y=0$ if and only if all $W_k$ are non-zero, which is the case when $1(W_1)1(W_2)\cdots 1(W_K)=1$. Otherwise, $Y$ is uniformly distributed in $\mathbb{F}_p \backslash \{0\}$, which is the case when $1(W_1)1(W_2)\cdots 1(W_K)=0$. Therefore, the scheme allows the server to securely compute $1(W_1)1(W_2)\cdots 1(W_K)$.

\noindent {\bf Phase II:} Since $\mathbb{F}_d^{\times}$ is isomorphic to $\mathbb{Z}_{d-1}$, the QSK-Prod of non-zero elements in $\mathbb{F}_d$ reduces to the QSK-Sum of elements in $\mathbb{Z}_{d-1}$. Let $\phi\colon \mathbb{F}_d^{\times} \rightarrow \mathbb{Z}_{d-1}$ be an isomorphism from $\mathbb{F}_{d}^{\times}$ to $\mathbb{Z}_{d-1}$ such that for any two elements $u,v \in \mathbb{F}_d^{\times}$, $\phi(u) + \phi(v) = \phi(uv)$. For non-zero $W_k$, define $w_k = \phi(W_k), \forall k \in [K]$. For $W_k=0$, define $w_k=\phi(\tilde{W}_k)$ where $\tilde{W}_k$ is generated uniformly over $\mathbb{F}_d^\times$ by User $k$. Then we apply the proposed QSK-Sum scheme over $\mathbb{Z}_{d-1}$ to compute $w_1+ w_2+ \cdots + w_K$ modulo-$(d-1)$, and by the isomorphism the secure computation of the  product of the $K$ non-zero elements is accomplished. Note that if any of the data symbols $W_k$ is zero, then this product is uniform over $\mathbb{F}_d^\times$ and independent of all data because of the way $\tilde{W}_k$ was generated. The cost for this phase is $(K/2) \log_d (d-1)$ qudits/computation. The idea of using the isomorphism  to compute the  product is not new, e.g., see \cite{jia2021SDBMM}. A key difference is that in  \cite{jia2021SDBMM}, the size of the additive group must be a prime to allow linear coding/decoding. This is not feasible in our setting as we require the additive group to have a size exactly equal to $d-1$ (otherwise the protocol would violate the security constraint). 

\noindent{\bf Security:} Note that Phase I reveals only the AND function, i.e., the product $1(W_1)1(W_2)\cdots 1(W_K)$, which is necessarily revealed by the desired product computation. If the AND is $0$, then because of the random choice of $\tilde{W}_k$ terms, the product computed in Phase II is independent and uniform over $\mathbb{F}_d^\times$, thus preserving security. If the AND is non-zero then by the security guarantee of the QSK-Sum protocol, Phase II reveals only the  desired product, so security is  preserved. 

\noindent {\bf Rate:} 
In Phase I, the cost is $(K/2)\log_d (p) \leq (K/2)\log_d (2K-1)$ qudits/computation, and in Phase II, the cost is $(K/2)\log_d (d-1)$ qudits/computation. Therefore, the rate achieved by the scheme is lower bounded by $ (2/K)/({\log_d(2K-1)+\log_d(d-1)})$ (computations/qudit). 

\subsubsection{Converse}
The converse proof is similar to the proof of the converse for Theorem \ref{thm:QSKSum} (QSK-Sum) in Section \ref{proof:converse_QSKSum}. It follows by replacing ${\bf 0}$ (the all $0$ vector) with ${\bf 1}$, the all $1$ vector of length $L$. The rest of the proof immediately applies as the server must be able to compute $W_k^{(L)}$ as well. $\hfill\blacksquare$

\begin{remark}
A classical scheme for product computation follows from the same 2 phase construction, using the FKN scheme \cite[Appendix B]{FKN} for Phase I, and additive secret sharing for Phase II. Thus Phase I incurs cost $K\log_d (p)\leq K\log_d(2K-1)$ dits/computation, and Phase II incurs cost $K\log_d(d-1)$. This produces an achievable classical rate $\frac{1/K}{\log_d(2K-1)+\log_d(d-1)}$. To our knowledge, a more efficient classical scheme is not available in the literature.
\end{remark}

\section{Conclusion}
The rate improvements found in this work are essentially due to (i) optimized batch-processing, (ii) the linearization, in Phase II, of the non-zero product to a sum via isomorphism, and (iii) exploiting linear combinations of inputs obtained through quantum entanglement via the $2$-sum protocol  \cite{song2020capacity,song2019capacitycollusion}. The latter advantage is similar to \emph{over-the-air} computation advantage in wireless, as noted in  \cite{Allaix_N_sum_box}. For the secure computation of \emph{arbitrary} functions in the PSQM setting, a promising approach suggested by this work for future efforts is to (i) find generalizations of the $2$-sum protocol, as explored in \cite{Allaix_N_sum_box}, and (ii)  find ways to linearize non-linear computations, e.g., by isomorphisms or by embedding  non-linear computations into linear computations, as explored in  \cite{zhao2021expand, QRPSM}. 

\bibliographystyle{IEEEtran}
\bibliography{Thesis.bib}

\end{document}